# Nanoscale chemical evolution of silicon negative electrodes characterized by low-loss STEM-EELS

*Maxime Boniface[1,2], Lucille Quazuguel[3], Julien Danet[1,2,‖], Dominique Guyomard[3], Philippe Moreau[3,\*], and Pascale Bayle-Guillemaud[1,2]*

[1] Université Grenoble Alpes, F-38054 Grenoble, France.

[2] CEA-INAC-MEM, F-38054 Grenoble, France.

[3] Institut des Matériaux Jean Rouxel (IMN), Université de Nantes, CNRS, 2 rue de la Houssinière, BP 32229, 44322 Nantes Cedex 3, France

\*Corresponding author: philippe.moreau@cnrs-imn.fr

‖**Present address: LCTS, CNRS, University of Bordeaux, Herakles-Safran, CEA, 3 Allee de La Boetie, F-33600 Pessac, France**

**Table of contents Graphic**

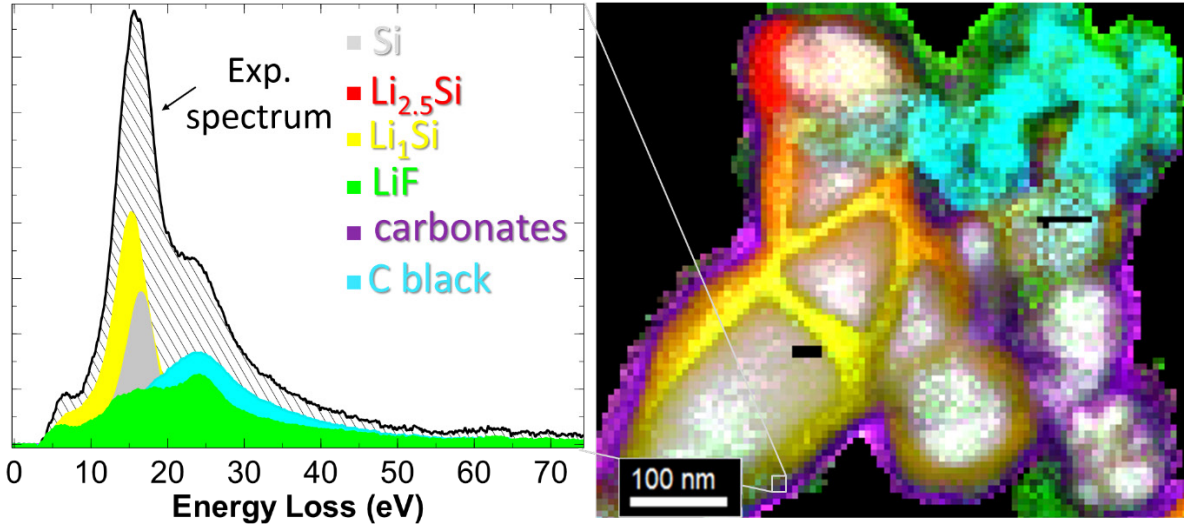

**Abstract:**

Continuous solid electrolyte interface (SEI) formation remains the limiting factor of the lifetime of silicon nanoparticles (SiNPs) based negative electrodes. Methods that could provide clear diagnosis of the electrode degradation are of utmost necessity to streamline further developments. We demonstrate that electron energy-loss spectroscopy (EELS) in a scanning transmission electron microscope (STEM) can be used to quickly map SEI components and quantify $Li_xSi$ alloys from single experiments, with resolutions down to 5 nm. Exploiting the low-loss part of the EEL spectrum allowed us to circumvent the degradation phenomena that have so far crippled the application of this technique on such beam-sensitive compounds. Our results provide unprecedented insight into silicon aging mechanisms in full cell configuration. We observe the morphology of the SEI to be extremely heterogeneous at the particle scale but with clear chemical evolutions with extended cycling coming from both SEI accumulation and a transition from lithium-rich carbonate-like compounds to lithium-poor ones. Thanks to the retrieval of several results from a single dataset, we were able to correlate local discrepancies in lithiation to the initial crystallinity of silicon as well as to the local SEI chemistry and morphology. This study emphasizes how initial heterogeneities in the percolating electronic network and the porosity affect SiNPs aggregates along cycling. These findings pinpoint the crucial role of an optimized formulation in silicon-based thick electrodes.

Silicon represents one of the most promising negative electrode materials for next generation lithium-ion batteries with a maximum uptake of 3.75 Li per Si atom, while graphite is limited to 1 Li atom per 6 C atoms[1–3] corresponding to a gravimetric capacity about ten times lower. Its use as a commercial electrode is however limited by the high irreversible capacity[4] caused by the colossal volume expansion (up to 270% for $Li_{3.75}Si$[5]) upon its electrochemical alloying reaction with lithium. This swelling causes interfaces to be repeatedly shifted during cycling, thus exposing unpassivated surfaces to the electrolyte and leading to continuous SEI formation[6] and lithium irreversible consumption. In addition to the loss of cyclable lithium in a full cell configuration[7], swelling also causes silicon nanoparticles (SiNPs) to be progressively isolated from both electronic[8,9] and ionic transport network because of the morphological instability and the SEI accumulation[10], respectively.

Averting these degradation phenomena is one of the major remaining challenges in the development of high energy-density Li-ion batteries. While controlling the size of nanoparticles[11,12] and cycling over limited capacities[13] have effectively mitigated the problems associated with pulverization, no such straightforward solution has been found regarding SEI instability, which is now the main hurdle towards using long-lived silicon-based electrodes in full cells. A fundamental understanding of the mechanisms of electrode degradation is required, and even more so to evaluate the strategies to be put forward to address them. Tremendous efforts have gone into developing a comprehensive model of SEI formation on silicon[7], with several studies concluding on a model similar to that of graphite[14,15] with the addition of lithium silicides[16]. Inorganic species related to salt anion decomposition, such as LiF when $LiPF_6$ is used, were found close to the electrode material while carbonates and polymeric species are situated closer to the electrolyte. Characterization of the SEI has so far mostly focused on describing its global chemical composition along cycling, with electrode-scale observations leading to suggestions on the nanoscale aging mechanism. Little is known about the actual SEI repartition in SiNPs aggregates that constitute the electrode. While NMR and FTIR provide a global view of the SEI chemistry[17–21], they offer no spatial information with respect to its structure or its morphology around SiNPs aggregates. XPS on the other hand can yield chemical spatially resolved information[7] although no better than at the micron scale and as a surface analysis technique, with limited depth profiling possibilities. Since the SEI varies at much smaller scale[16,22–24], developing a comprehensive picture of the SEI characteristics, such as its own microstructure and disposition around SiNPs and across the electrode, and their relationship to nanoparticles lithiation mechanisms along cycling requires chemical information at the nanoscale.

Scanning transmission electron microscopy (STEM) in a transmission electron microscope (TEM) combined with electron energy-loss spectroscopy (EELS) conveniently provides the spatial resolution that can be less than 0.1 nm in modern TEMs, with chemical and atomic bonding investigation. EELS ionization edges are known to contain quantitative information on both elemental composition and chemical environment, which has allowed numerous high-resolution studies on positive[25–27] and negative electrode materials[28]. *Operando* experiments have provided unprecedented insight into fundamental lithiation processes[29,30]. Their success depends on the design of model systems, which include unconventional electrolytes and customized geometries of the cell. As such, findings cannot yield a direct understanding of the failure mechanisms at work in real-scale of full battery systems. *Ex situ* measurements thus definitely remain relevant to study SEI-related aging, the morphology of the active material in slurry-based electrodes as well as their evolution along cycling. Applying STEM/EELS to negative electrode materials and SEI compounds has however been proven particularly challenging due to the extreme electron beam-sensitive nature of these compounds. Consequently, there are currently no published results of dose-controlled, nanometer-resolved SEI studies on silicon.

In this work, we demonstrate how this can be achieved through the study of the low-energy part of the EEL spectrum (<50 eV), called plasmon losses, as it was previously demonstrated for organic species[31,32], silicon-based compounds[33] or phase determination in cycled LiFePO$_4$[34–36]. Plasmon peaks are considerably more intense than high energy core-loss edges [37] at similar electron doses. Decent signal-noise ratio (SNR) acquisitions can thus be obtained within timeframes compatible with minimal sample degradation providing optimized experimental conditions are used. These advanced acquisition and analysis protocols allow us to map SEI species and Li$_x$Si alloys with resolutions down to 5 nanometers. Preserving then all the information contained in the SEI at the aggregate and single nanoparticle level, we put forward its inhomogeneity and could establish how SEI affects the lithiation process in the electrode. Moreover, a precise determination of the plasmon energy of silicon giving an indication on its crystallinity, provides us with clues on the cycling history of each nanoparticle inside aggregates.

Silicon-based electrodes were prepared using an aqueous slurry of crystalline nanoparticles (80wt%, Nanostructured & Amorphous Materials), carboxymethylcellulose (CMC, 8wt%, Sigma-Aldrich) and carbon black (12wt%, superP) to obtain mass loadings of 1±0.1mg.cm$^{-2}$. Cycling was performed in Swagelok® cells, with 1M LiPF$_6$ in EC:DEC carbonate electrolyte with 10wt% FEC additive, in full battery configuration against Li(Ni$_{1/3}$Mn$_{1/3}$Co$_{1/3}$)O$_2$. The capacity was limited to 1/3$^{rd}$ of the silicon electrode capacity (1200 mAh.g$^{-1}$) at a rate of 480 mA.g$^{-1}$ (1 Li per Si atom in 2 hours). Results of the galvanostatic cycling were published in a previous study[7]. They indicate that the initial capacity of the Li-ion cell is drastically decreased to approximately 300 mAh.g$^{-1}$ after 100 cycles, hence a low capacity retention that need to be understood. After disassembly, the electrodes were not rinsed in order to prevent any risk of dissolving species present in the SEI, and left to dry in an argon-filled glovebox (<1ppm H$_2$O, <1ppm O$_2$), then gently scratched over TEM copper mesh grids to transfer

SiNP aggregates. These grids were subsequently mounted onto a Gatan vacuum transfer holder to directly bring the sample from the glovebox to the microscope without any exposure to oxygen or moisture.

Most species present at the negative electrode of Li-ion batteries after cycling show high contents of light elements such as lithium, carbon, oxygen and fluorine, which have sub-1keV energy-loss edges that are well suited for EELS. However, typical SEI compounds such as $Li_2CO_3$ show critical doses as low as 750e$^-$/Å², which resemble those found for biological samples[38]. Experimental conditions associated with the study of edges inevitably cause the loss of the characteristic spectral features that can be used to identify the chemical environment of elements (such as C-O and C=O bonds), so that only elemental maps can be obtained[39,40]. Further exposure leads to mass loss through radiolysis in the SEI and lithium surface migration through knock-on damage in lithium-silicon alloys[41] (Figures S2). One successful approach has been to cool down samples to $LN_2$ temperatures[42], which is known to increase the critical dose by up to 3 orders of magnitude[37]. But doing so on cycled negative electrodes materials, which also need to be protected from $O_2$ and $H_2O$, requires specific equipment, low $H_2O$ content in the TEM vacuum and also limited drift during mapping. In addition to increased overall experimental time, this low temperature technique is difficult to set-up and unlikely to be disseminated.

Although plasmons cannot usually be used for quantitative elemental analysis, their shapes vary substantially between specimens and provide an additional characteristic fingerprint of compounds[31]. This can be used to identify SEI species, $Li_xSi$ alloys and other electrode components. Acquiring decent plasmons, while keeping doses low enough, is however by no means straightforward. Experimental parameters should indeed be chosen very carefully to both optimize the SNR and mitigate the materials degradation during 2D mapping. The fact that there is an incompressible readout time (~3-5ms with our setup) during which the sample is still exposed makes a low probe current desirable. For this, probe currents of no more than 6pA were used during this study. The collection (β) and convergence (α) semi-angles are usually chosen following the guidelines that β should be about twice the characteristic scattering angle and that the convergence angle α should not exceed β/2. Taking a mean energy of 20eV for plasmons, this would correspond to $β ≈ 2θ_E = 0.12$mrad. Using effectively low α values would result in an extremely large probe incompatible with STEM measurements. In practice, we chose α so as to obtain an electron probe that is adapted to the step size associated with an acceptable electron dose for our compounds, and deduce the best β value so as to have an optimized SNR. Aiming for a dose of 300e$^-$/Å² and given a probe current of 6pA and a dwell time of 20ms (including readout), the optimum corresponds to a step size of 5nm. Considering the spatial distribution of our probe, a FWHM of 1.5nm will correspond to an acceptable ~5% of the beam overlapping on adjacent pixels. Using very low convergence angles has the additional advantage of widening the electron beam, which further reduces the areal electron dose. By controlling the electron dose this way, we are able to conduct consecutive measurements with negligible degradation (Figure S3), thus avoiding both radiolysis and sputtering phenomena - exemplified in the Supporting Information.

STEM-EELS analyses were performed in a FEI Tecnai Osiris operated at 200kV equipped with a Quantum SE Gatan Imaging Filter (GIF). An energy resolution of 1eV was routinely obtained at a dispersion of 0.1eV/channel, with β≈5mrad and α≈2.5mrad. Spectrum images (SIs) were recorded by scanning the beam across the specimen and recording EEL spectra at each position. Since both EELS acquisition and readout times are limited to a minimum (using the highest possible vertical binning before saturation), SIs are swiftly acquired (<20ms per voxel including spatial drift correction), which allows us to analyze large particle clusters (~1μm²) in short timeframes (below 10 minutes). After energy alignment, dispersion calibration and Fourier-log deconvolution routines, principal component analysis[43] was applied as a noise-removing method. In order to interpret the resulting datasets, a database of the reference spectra from known electrode components and SEI compounds (Figure 1a) as well as Li-Si alloys (Figure 1c) first needs to be constituted[44]. This was done through analyses of reference samples of the compounds of interest. At high resolution, the electron dose increases and experimental spectra inevitably show shifts or dampening of certain features compared to reference ones. To circumvent this, reference spectra were acquired over a range of different doses (Figure 1b) for major SEI compounds, such as LiF and $Li_2CO_3$, so that radiolysis could be anticipated in experimental spectra (Figure 1d). The contribution of individual reference spectra to an experimental spectrum could then be calculated through the Multiple Linear Least Square (MLLS) fitting routine between 0 and 70eV (Figure 1d). Iterating this protocol over the entire SIs allowed us to reliably map each species relative quantity at the particle scale. These maps can then be combined into RGB images showcasing local nanostructures. Spectra from thick areas and surface plasmons that had a poor SNR were filtered out. It is also important to note that since $Li_2CO_3$ is the only carbonate that was considered in this fitting routine, its signal most likely includes the contributions of the whole array of alkyl carbonates such as lithium ethyl dicarbonate (LiEDC)[45] that are thought to account for a large part of the SEI. We assume that these larger carbonates give similar spectra.

As a typical example of the MLLS fit, the mapping in Figure 1e illustrates the output of this method on particles from a SiNP-based electrode after its first lithiation (1$^{st}$ charge of the full battery) to 3000mAh.g$^{-1}$ (80% of the theoretical full capacity), with an 8nm step size. More information regarding the fitting process of overlapping compounds as well as the components presented separately can be found in the Supporting Information. The silicon (white) and $Li_xSi$ components (red and yellow) show well-known core-shell structures as well as fully lithiated smaller nanoparticles. The SEI can be seen in the form of particle-like LiF (green) deposits that can be almost 100 nm across, as well as a thinner, more continuous layer of carbonates (purple). Such morphology for LiF was indirectly observed in previous work and appears here in plain view[46–48]. Lithiation is observed to proceed in a very heterogeneous manner on the aggregate scale, with different $Li_xSi$ alloys observed on adjacent SiNPs, whereas homogeneous compositions are observed in individual nanoparticles.

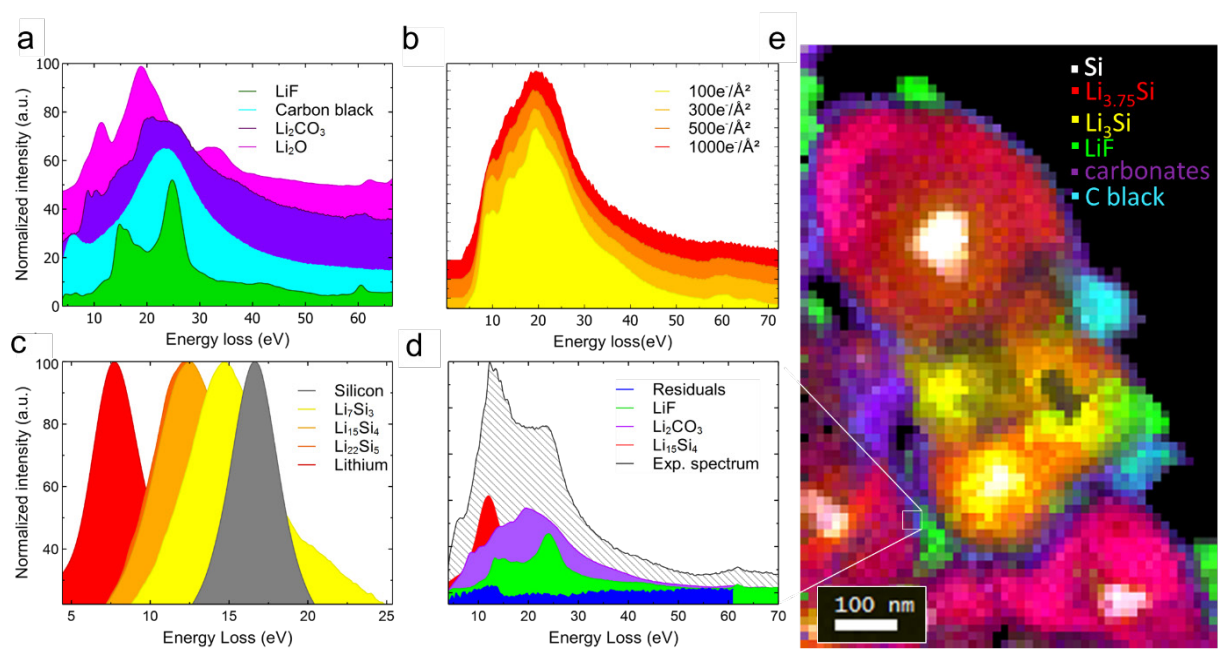

**Figure 1.** a) EELS reference plasmon spectra for LiF, carbon black, $Li_2CO_3$, and $Li_2O$. b) Plasmon spectra of $Li_2CO_3$ acquired at different electron doses. c) Reference plasmon spectra of $Li_xSi$ alloys. d) Example of MLLS fitting on the experimental spectrum integrated over 4 pixels of (e). e) Color map of LiF(green), $Li_2CO_3$(purple), carbon black(teal), $Li_7Si_3$(yellow), $Li_{15}Si_4$(red) and Si(white), obtained through MLLS fitting of a spectrum image from an electrode charged to 3000mAh.g$^{-1}$. The scale bar is 100 nm.

Complementarily, it is possible to go beyond the MLLS method reconstruction regarding the interpretation of plasmons. For SEI compounds, low-loss spectra are often complex to interpret and this "finger print" based method must be preferred. On the other hand, Si and $Li_xSi$ alloys can be reasonably approximated to behave like free-electron metals and, as such, their low-loss spectra can be fitted with the simple Drude model[37,49] (eq.1). This model describes the energy-loss function Im (-1/ ε), where ε is the complex permittivity of the probed material, Γ is a damping factor related to the full width at half maximum (FWHM) of the peak, and $E_p$ the bulk plasmon energy.

$$Im\left[\frac{1}{\varepsilon(E)}\right] = \frac{E\Gamma E_p^2}{\left[(E^2-E_p^2)^2+(E\Gamma)^2\right]} \text{ (eq.1)}$$

The determination of $E_p$ allowed us to go further than the MLLS decomposition process to map lithium-silicon alloys. Since $E_p$ decreases continuously with increasing lithium content, an empirical quantification method can be extrapolated from the analysis of reference samples with known stochiometries[44]. Applying this method across spectrum images produced quantitative $Li_xSi$ composition maps (Figure S4). It also provides an opportunity to evaluate the crystallinity of silicon in delithiated samples by exploiting the ~0.1eV plasmon peak energy difference between crystalline and amorphous silicon[50].

Our ability to map the different components of these composite nanostructures is a very appropriate way to explore at the nanometer scale electrode failure mechanisms with a more systemic approach along cycling. Analyses were performed on electrodes at their 1st and 100th charge and a representative excerpt of these results is reported in Figure 2. After the 1st charge (lithiation of silicon), numerous acquisitions showed nanoparticles in their pristine state without any evidence of lithiation, yet they were covered by large amounts of LiF (Figure 2a). One would assume SEI-covered particles to be active in the overall redox charging process and show some degree of lithiation. This suggests that LiF deposition could happen through $LiPF_6$ chemical decomposition, which has been suggested in previous studies[51,52], in addition to FEC reduction[53–55,48]. This is further exemplified in Figure 2c, which features a LiF matrix engulfing nanoparticles after 100 cycles. This matrix is unlikely to have formed electrochemically given the poor conductivity of LiF. Such large agglomerates are expected to considerably hinder Li$^+$ transport in the electrodes through pore clogging[19,56]. LiF is however not the sole responsible of the accumulation of the SEI, as evidenced by the large carbonate layers (~100nm) that were observed in electrodes after 100 cycles (Figure 2d). Further analysis of the low-loss spectra of this layer, whose plasmon was already identified as carbonates through MLLS mapping, reveals a diminished Li-K edge intensity (Figure 2e) compared to its counterpart at the 1$^{st}$ cycle, indicating a chemical evolution towards compounds with lower Li contents after extended cycling. This supports the aging mechanism proposed in previous work where the potential of end of charge of the silicon electrode slips toward less reductive values as cycling progresses[7,18], for which lithium-rich SEI compounds are unlikely to form[57]. During the first cycles (Figure 2b), lithiated parts of the electrodes systematically showed SEIs that followed almost the same pattern: moderate particle-like LiF formation, mostly situated in the innermost part of the SEI as it has been suggested in other works; a more conformal carbonate layer, with thicknesses in the range of 15 to 25 nm. This is coherent with previous reports[51,58–60] and was made possible thanks to the negligible mass loss of these compounds under the electron beam with our setup. Although traces of Li$_2$O could be detected in a few experiments, the presence of this compound is thought to be anecdotal and lithium oxide cannot be considered one of the main SEI component, in accordance with previous reports[22,46].

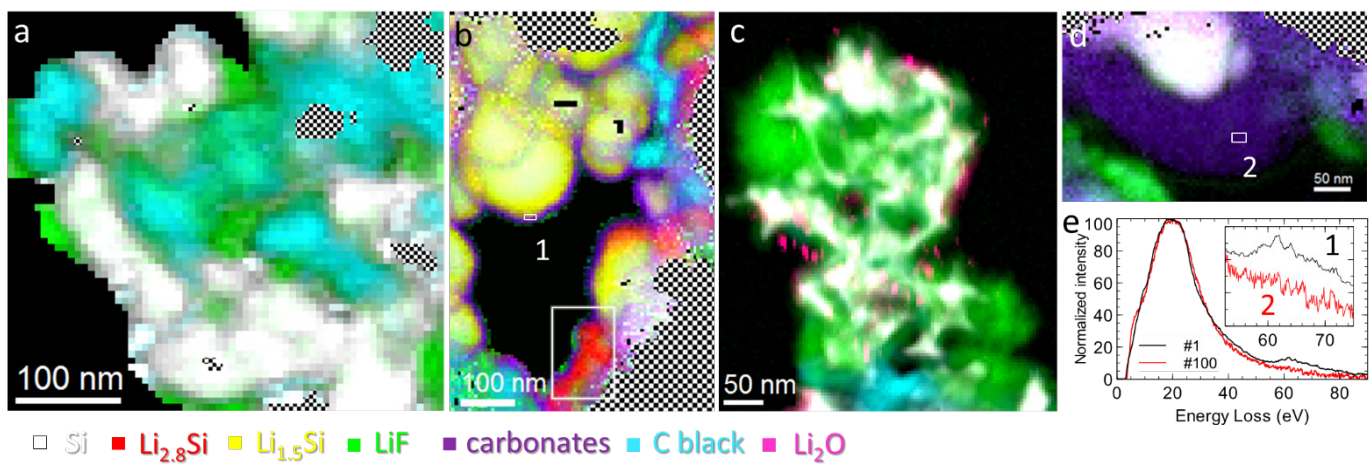

□ Si  ■ Li$_{2.8}$Si  ■ Li$_{1.5}$Si  ■ LiF  ■ carbonates  ■ C black  ■ Li$_2$O

**Figure 2.** Color maps performed on Si electrodes cycled at 30% of their theoretical capacity (i.e. 1200 mAh.g$^{-1}$). LiF (green), carbonates (purple), silicon (white), Li$_x$Si (yellow-red), carbon black (teal) and Li$_2$O (pink). The checkered areas correspond to regions where the thickness is too large to allow a relevant signal decomposition a) After the 1$^{st}$ charge of the Li-ion cell (1$^{st}$ lithiation of Si), unlithiated, pristine silicon nanoparticles covered by large amounts of LiF. b) After the 1$^{st}$ charge of the Li-ion cell (1$^{st}$ lithiation of Si), c-Si/Li$_x$Si core-shells covered by ~20nm thick layers of carbonates and particle-like LiF formation as well as smaller, fully lithiated, highlighted nanoparticles (in white square).  c) After 100$^{th}$ discharge of the Li-ion cell (lithiated Si electrode), delithiated nanoparticles trapped in a LiF matrix-like deposit. d) After 100$^{th}$ discharge, thick (~100nm) carbonate layer on a single nanoparticle. e) Low-loss EEL spectra acquired on carbonates observed at the 1$^{st}$ and 100$^{th}$ cycles (marked respectively areas 1 and 2) showing a diminished lithium content (Li K-edge around 60 eV) for carbonates formed after extended cycling.

Even at limited capacity (30% of SOC), the smallest nanoparticles, often joined together, (Figure 2b, highlighted area) seem to be lithiated to their core, in contrast to larger ones. This apparent difference in lithiation mechanisms necessitates further analysis of the silicon spectrum in nanoparticles of different sizes. In order to do so, we studied delithiated nanoparticles aggregates from discharged batteries. It should be first mentioned that E$_p$ values at the surface (< 5 nm from surface) should be taken with care since those fits are quite sensitive to surface plasmons[37] and are convoluted with spectra from silicon oxides and the SEI that all have higher plasmon bulk energies. This results in surface artifacts in Figure 3a, where the extreme surface always appears red to yellow (approx. 16.75 eV), which wrongly conveys the idea of crystalline silicon at the interface, whereas more accurate values around 16.55 eV (dark blue to green areas eg amorphous silicon) were fitted on the silicon plasmons a few nm inside the aggregates. Delithiated particle aggregates showed extremely different morphologies for different SiNPs populations. HRTEM images of the two SiNPs populations present in our sample are shown in Figure 3c and 3d. While lithiation proceeds in a core-shell fashion in large (80-150nm) nanoparticles (Figure 3a), with fitted plasmon energies around 16.7-16.85eV, corresponding to crystalline silicon, the SiNPs of our second smaller population show sintering leading to large amorphous domains, several hundredths of nanometers across (Figure 3b). These domains are much larger than individual particles, usually in the 30-50nm range (Figure 3d). However, size alone cannot be responsible for these different behaviors, since it has been shown that shell growth occurs similarly regardless of nanoparticle absolute diameter[11]. This behavior upon cycling appears to be the result of preferential lithiation paths in these smaller SiNPs, which HRTEM images reveal to be polycrystalline in their pristine state (Figure 3d). Grain boundaries will in fact act as fast $Li^+$ diffusion paths[40,61] during the charge (Si electrode lithiation), as previously shown with crystalline defects in Si microparticles[62]. This is further illustrated in the nanoparticle highlighted in Figure 4b, with the formation of two secondary core-shell structures separated by a lithiated boundary (probably initially a defect across the particle) with concentrations of 0.5-1 Li per Si atom.

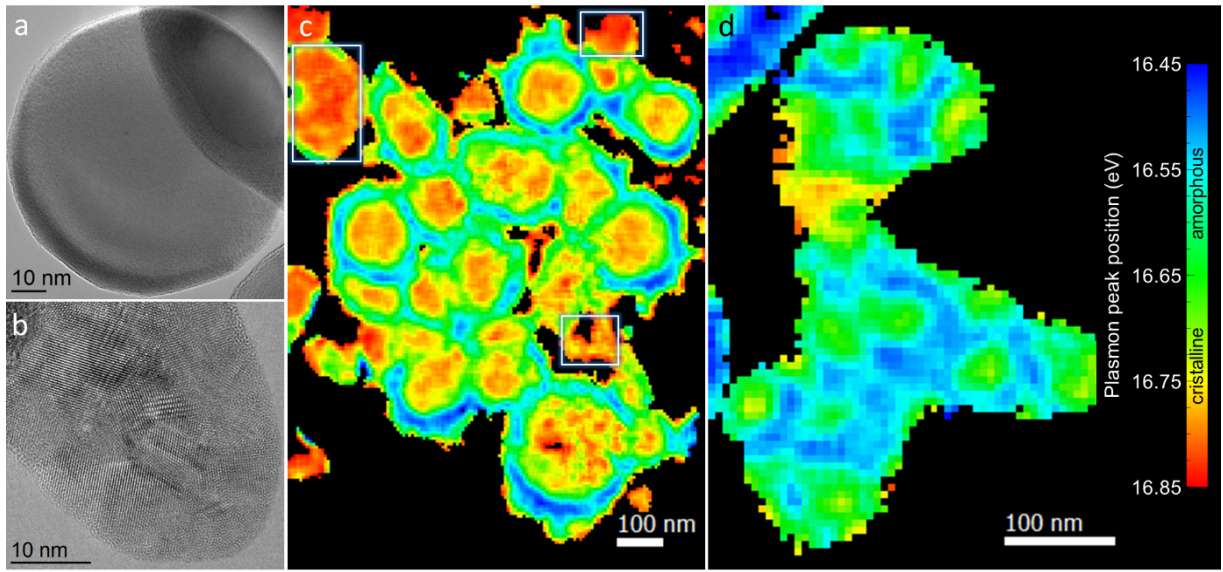

**Figure 3:** Mappings of plasmon peak maximum positions obtained from fitting with the simplified Drude model (eq. 1) on SIs acquired on SiNPs from delithiated electrodes and HRTEM images of the 2 corresponding SiNPs populations. a) HRTEM image of a monocrystalline particle typical of the first SiNP population. b) HRTEM image of a polycrystalline particle typical of the second SiNP population. c) Large c-Si/a-Si core shell structures resulting from the delithiation of the monocrystalline SiNP population illustrated in a) after 100 cycles. Pristine (unreacted) crystalline SiNPs are highlighted by white squares. d) Amorphous silicon matrix with remaining crystalline domains resulting from the delithiation of the polycrystalline particle population illustrated in b) after 10 cycles.

Regardless of initial crystallinity, inhomogeneities in lithium content were observed among SiNPs aggregates as well as individual nanoparticles. We observed a continuous range of $E_p$ values revealing variable Li-Si alloy compositions in charged particles. In delithiated aggregates, plasmons energies are also observed to vary continuously between 16.85 to 16.45 eV. This corresponds to the existence of intermediate orderings rather than a sharp boundary between crystalline and amorphous domains This is thought to be related to the presence of varying quantities of silicon clusters in amorphous silicon after delithiation,[63] depending on the depth of the previous charge[64,65]. Considering this, the lithiation history of individual nanoparticles can be followed by mapping $E_p$. The gradient of $E_p$ values in the shells of the delithiated SiNPs aggregate of Figure 3a thus demonstrates disparities in lithiation not only between adjacent nanoparticles, but also across each shell.

This inhomogeneity was directly observed in Figure 4b, with varying lithium concentrations from 0.5 to 1.8 lithium per silicon atom measured in the shell of the highlighted nanoparticle (in the white square). This indicates that lithiation kinetics can be diffusion-controlled, with the reaction proceeding preferentially near the edge of porosities were Li$^+$ ions are most readily available, and particles situated on the inside of aggregates, only accessible by intra-SiNPs diffusion[66], exhibiting lower lithium contents. Differences in lithium content also appear to be closely correlated with the local SEI composition. This is demonstrated in Figure 4. In Figure 4a, the SEI composition is obtained from the MLLS method. In Figure 4b of exactly the same area, the analysis of $E_p$ values provides the precise composition of the alloys. In Figure 4c, the detailed analysis of $E_p$ for unlithiated Si is given. Based on the comparison of Figure 4a and b, particles near green, LiF-rich areas are shown to present lithiated shells of a wide range of alloy compositions. SiNPs in the purple corresponding to more carbonate-rich area, on the contrary, remain delithiated. The mapping of amorphous Si areas (Figure 4c) reveals that the particles that remained delithiated have developed the core-shell structure associated with previous cycles. This suggests that these SiNPs were initially active then were subsequently excluded from cycling via carbonate buildups, in which Li$^+$ ions diffuse poorly[67]. SEI accumulation with cycling makes the lithiation of these SiNPs increasingly energetically unfavorable and results in the loss of available active matter[7,10,18]. Particles were also observed to disconnect from the electronic percolating network along cycling. $E_p$ mappings of charged Li-ion cells (lithiated Si electrodes) show large aggregates of unlithiated

SiNPs, while conversely a few lithiated Si particles were found in delithiated electrodes (Figure S5). The overall behavior is coherent with NMR results from previous work[7,10], namely that on the electrode scale lithiated alloys quantities are negligible in discharged batteries. Furthermore, particles were observed in their pristine, crystalline state even in electrodes at their 100$^{th}$ cycle and in spite of their close proximity with active particles (Figure 3a and S4-b and c). Volume changes associated with lithiation indeed make it exceedingly difficult to preserve electronic percolating networks in SiNPs negative electrodes.

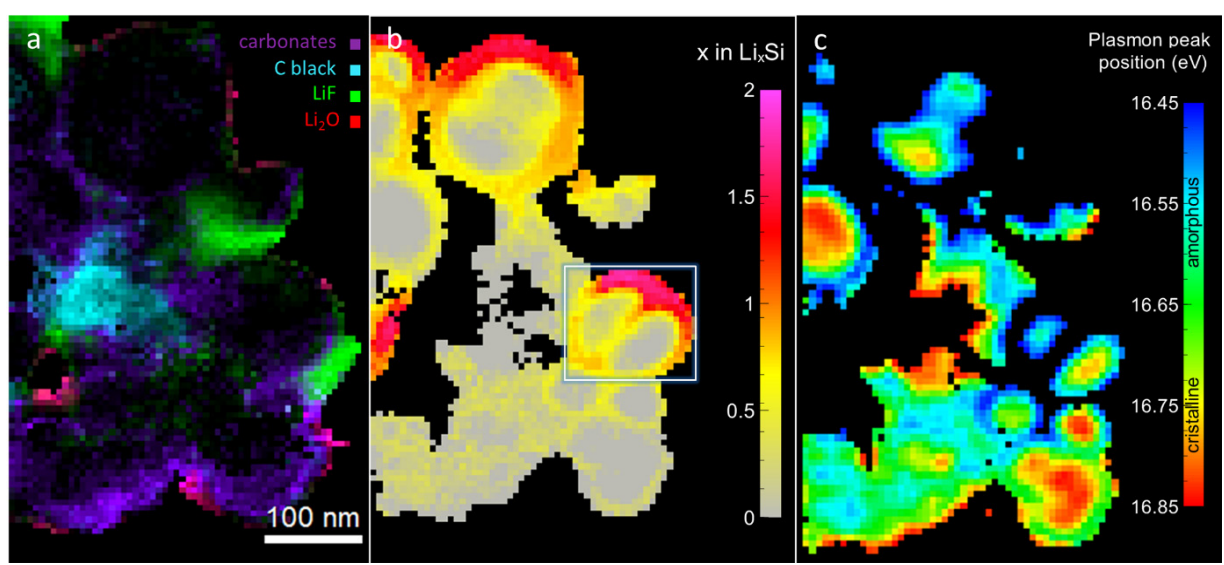

**Figure 4.** SEI mapping, Li$_x$Si alloy composition mapping and Si crystallinity mapping of a SiNP aggregate from an electrode after its 10$^{th}$ lithiation to 1200mAh.g$^{-1}$ showing local SEI composition and lithiation correlation. a) SEI map showing inhomogeneous LiF (green), carbonate (purple) and Li$_2$O (red) deposits, as well as carbon black (teal). b) Li$_x$Si alloys composition map showing a preferential lithiation lithium diffusion path along the grain boundary of the highlighted SiNP. Particles in SEI-poor or LiF-rich areas appear more likely to be lithiated. c) Plasmon peak position mapping corresponding to silicon degree of crystallinity.

In conclusion, we developed low-loss EELS mapping into a powerful diagnostic tool for aging analysis of lithium-ion battery silicon negative electrodes. Innovative acquisition protocols allowed us to gather exploitable chemical information on the SEI at extremely low electron doses, which translates to the ability to study the evolution of the SEI along cycling at an unprecedented spatial resolution. Furthermore, we show that the analyses of the low-loss spectra can be refined to yield alloy composition and an estimation of the crystallinity of silicon nanoparticles in cycled electrodes. Our method allows us to exploit a single low-loss dataset for diverse but highly complementary information to reach an exhaustive assessment of the local state of health of electrodes. This was successfully exploited to reveal electrode aging mechanisms at the particle scale, with direct observations of SiNP disconnection from both electronic and ionic conduction pathways and crystallinity-dependent lithiation mechanisms. Findings suggest that capacity retention could be improved through both a better control of the porosity and mixing of carbon black to mitigate those disconnection phenomena. Furthermore, we took advantage of the use of a bimodal Si nanopowder, with mono- and poly-crystalline particles, to evaluate the difference in lithiation mechanism. Our study demonstrates how detrimental such dispersion is, creating local electrochemical potential gradients and a strong heterogeneous lithiation process of the electrode. The fact that some SiNPs have to become overlithiated compared to the $1/3^{rd}$ capacity of the electrode prevents the mitigation of the swelling. All the benefits of the limited capacity are thus lost: the differential swelling increases constrains in the whole electrode and the pulverization is amplified and leads to enhanced SEI formation. In addition, in case of large carbonate-based SEI growth, embedded SiNPs become electrochemically inactive. The phenomenon also implies a preferential lithiation of some more accessible particles, hence a further heterogeneous electrode lithiation. These results point out the special care to be paid to the nanoscale homogeneity of the electrical wiring and accessibility of porosity in these highly electrolyte-reactive electrodes.

The tremendous spatial resolution of STEM-EELS can now be taken advantage of to compare SEIs from different electrolyte formulations or study commercially relevant alloying negative electrode materials such as Si-based intermetallics, SiO composite particles, or pre-lithiated silicon. There is still much room for improvement, both in lowering the electron dose and interpreting plasmons, which makes applying this methodology all the more promising. Our protocol could be extended to other beam-sensitive composite nanostructures on which core-loss EELS has been difficult so far, with the only criterion being that domains stay larger than the minimum pixel size associated with the smallest dose required for acceptable plasmon acquisition.

## ASSOCIATED CONTENT
**Supporting information**
Description of the material.

Figure S1: Morphology of individual phases revealed through MLLS mapping.

Figure S2: Examples of radiolysis and sputtering damage after prolonged exposition to the electron beam.

Figure S3: Consecutive measurements with an optimized protocol showing negligible degradation.

Figure S4: Quantification method to determine the lithium content in $Li_xSi$ alloys

Figure S5: Delithiated particles in charged electrodes & lithium trapped in discharged electrodes

This material is available free of charge via the Internet at http://pubs.acs.org

## AUTHOR INFORMATION
**Corresponding Author**

*E-mail: philippe.moreau@cnrs-imn.fr
**Notes**
The authors declare no competing financial interest.

**ACKNOWLEDGEMENTS**

This research was funded by the European Research Council (ERC), EU FP7 Energy.2013.7.3.3 program on Understanding interfaces in rechargeable batteries and super-capacitors through in situ methods, Grant award No. 608491 on project "Battery and Supercapacitors Characterization and Testing" (BACCARA).